# Room-Temperature Polariton Lasing from CdSe core-only Nanoplatelets

*Francisco Freire-Fernández, [†,1] Nathan G. Sinai,[†,1] Max J. H. Tan, [1] Sang-Min Park,[1] Eric Koessler, [2] Todd D. Krauss,*[,2] Pengfei Huo*[,2] and Teri W. Odom*[,1]*

[1] Department of Chemistry, Northwestern University, Evanston, IL 60208; [2] Department of Chemistry, University of Rochester, Rochester, NY 14627

[†]These authors contributed equally.

**Abstract**

This paper reports how CdSe core-only nanoplatelets coupled with plasmonic Al nanoparticle lattices can exhibit exciton-polariton lasing. By improving a procedure to synthesize monodisperse 4-monolayer CdSe nanoplatelets, we could resolve polariton decay dynamics and pathways. Experiment and theory confirmed that the system is in the strong coupling regime based on anti-crossings in the dispersion diagrams and magnitude of the Rabi splitting values. Notably, polariton lasing is observed only for cavity lattice periodicities that exhibit specific dispersive characteristics that enable polariton accumulation. The threshold of polariton lasing is 25-fold lower than reported photon lasing values from CdSe nanoplatelets in similar cavity designs. This open-cavity platform offers a simple approach to control exciton polaritons anticipated to benefit quantum information processing, optoelectronics, and chemical reactions.

**TOC GRAPHIC**

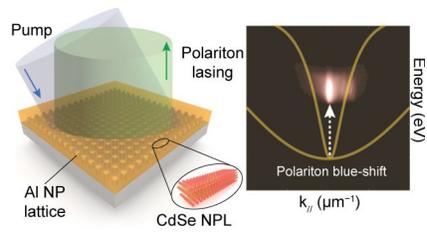

Strong light-matter coupling and the generation of exciton-polaritons are attracting attention in quantum information processing,[1,2] optical and spin-based electronic devices,[3,4] and controlling chemical reactivity.[5,6] Polaritons are hybrid light-matter states that form when the interaction strength between an optical cavity and emitter exceeds their individual losses. In a strongly coupled system, energy is coherently exchanged between the electronic state of the emitter and the optical mode of the cavity.[7,8] Recently, polaritons have been pursued to understand mechanisms of light generation since, unlike photon lasing, polariton lasing does not depend on the population inversion of excited states but on the build-up of polariton population in the same quantum state.[9,10] Since the coherent emission of photons by radiative decay of polaritons occurs spontaneously, polariton lasing occurs at considerably lower thresholds.[9,11] This distinct property has resulted in prospects to realize electrically-driven, room temperature polariton lasing by optimizing both gain material and cavity design.[12-15]

Semiconducting CdSe nanoplatelets (NPLs) are promising exciton materials for strong coupling because of their out-of-plane quantum confinement in 1D and large in-plane oscillator strength.[16,17] Compared to other emitters used in strongly coupled systems (e.g., organic dyes,[18] emissive metal-organic frameworks,[19] and perovskites[15,20,21]), the crystal structure of CdSe NPLs results in increased chemical stability;[22] moreover, their exciton states can be n- or p-doped under an applied voltage.[23] CdSe NPL excitons can be delocalized across their entire 2D area in the absence of defects, which minimizes non-radiative optical losses from biexciton Auger recombination;[24,25] however, defects can lead to exciton localization.[26] These features have enabled amplified spontaneous emission as well as photon lasing under pulsed[27,28] and continuous-wave pump excitation.[29]

The primary approach to reduce thresholds using CdSe NPLs as gain material has focused on core/shell engineering or lateral size modulation to mitigate Auger-based losses.[24, 25] However, polariton lasing offers an alternative route to realize efficient coherent light sources. Previous work has demonstrated strong coupling using core-only CdSe NPLs in Fabry-Pérot,[30] gold film,[31] and nanohole array[32] cavities, but polariton lasing has not been reported. Plasmonic nanoparticle lattices are attractive cavities for polariton lasing given that they are a well-established cavity for photon lasing; their large scattering cross-sections from the localized surface plasmons of nanoparticles combined with the in-plane diffraction (Rayleigh anomaly) modes[33-35] form surface lattice resonances (SLRs) that support distributed feedback.[36-41] In addition, SLR modes exhibit polarization-dependent dispersion properties.[33, 35] Importantly, strong coupling using these cavities with organic dyes,[18, 42] metal-organic frameworks,[19] and 2D perovskites,[20] have been reported where the open-cavity architecture of plasmonic lattices enabled precise tuning over the degree of coupling.

Here we show room-temperature polariton lasing from CdSe NPL films strongly coupled to plasmonic Al nanoparticle lattices. By optimizing the synthesis and purification protocol to produce monodisperse 4-monolayer CdSe NPLs, we could differentiate polariton decay dynamics from that of common CdSe impurities. We discovered that although strong coupling was supported by different lattice periodicities, polariton lasing was only observed at specific exciton-cavity detuning. Angle-resolved transmission measurements exhibited anti-crossing and band-bending signatures that closely matched Jaynes-Cummings-Hopfield calculations; the fitted Rabi splitting energies were in the strong coupling regime. When the devices were optically pumped, lasing from the lower polariton (LP) band was observed at a threshold of ~6 $\mu J/cm^2$, 25 times lower than values reported for photon lasing. The polaritonic nature of the emission was supported by static and time-resolved photoluminescence measurements below lasing threshold and Holstein-Tavis-Cummings calculations.

**RESULTS AND DISCUSSION**

**Figure 1a** depicts the light and matter components for a strongly coupled system based on an Al plasmonic nanoparticle lattice coated with a CdSe NPL thin film. Al nanoparticles (height $h \approx$ 55 nm, diameter $d \approx$ 95 nm) were patterned by electron-beam lithography on high-refractive index ($n$) glass substrates ($n_{glass} \approx 1.578$ @ 500 nm) to reduce refractive index differences with the CdSe NPL films ($n_{film} \approx 1.73$ @ 475 nm, thickness $t = 95$ nm) that can result in hybrid optical modes,[43] including waveguide-SLRs[32, 38, 44] (**Figure S1**) with more complicated features. Synthesis of 4-

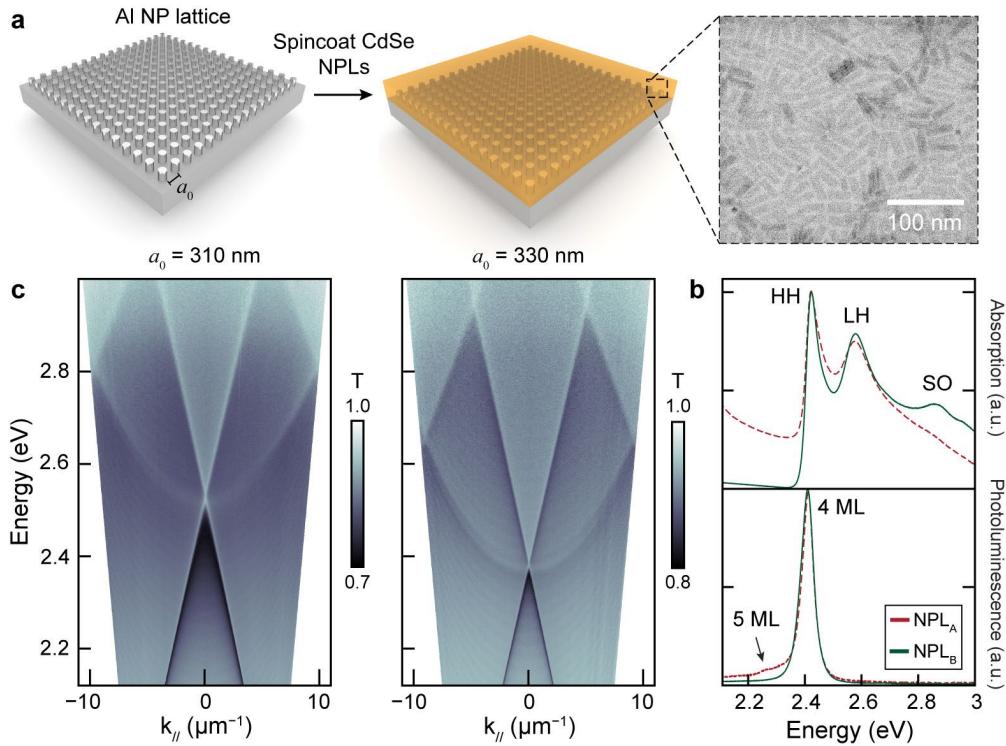

**Figure 1: Strongly coupled CdSe NPL-Al nanoparticle lattice system.** (a) Scheme of components. Inset shows TEM image of CdSe $NPL_B$. (b) Absorption and photoluminescence spectra of $NPL_A$ and $NPL_B$ batches. (c) TE-polarized optical transmission dispersion diagrams of bare Al nanoparticle lattices with different periodicities, $a_0$.

monolayer (ML) CdSe NPLs generally followed reported protocols,[27, 45, 46] but we modified and scaled the procedure to achieve monodisperse 4-ML CdSe NPLs with high brightness (**Supporting Information: S1**). Transmission electron microscopy of NPLs from our optimized protocol shows uniform lateral sizes of ca. 10 nm by 30 nm (Figure 1a). Although the synthesis of CdSe NPLs has been widely adopted,[27, 29, 45, 47, 48] we discovered that the quality of products depended critically on the initial degassing of the reagents. We speculate that this step involves the formation of CdSe seeds that can then direct NPL growth. For comparison, we synthesized and purified two CdSe NPL batches under the same conditions except for the degassing step: (1) at 80 °C for 1 h according to literature[27] (NPL$_A$); and (2) at 95° C for 30 min in our modified procedure (NPL$_B$). Films were produced by concentrating NPL solutions to 60 mg/mL in cyclohexane and then spin-casting at 4000 rpm on transparent substrates (**Supporting Information: S2**).

**Figure 1b** shows both normalized absorption and photoluminescence of CdSe NPL$_A$ and NPL$_B$ films. In both batches, the absorption spectra had two distinct bands (2.42 eV and 2.58 eV) that can be assigned as heavy hole (HH) and light hole (LH) excitons, respectively. However, the spin-orbit split-off band (SO) at 2.86 eV was exclusively detected in NPL$_B$, and CdSe NPL$_A$ exhibited a broad absorption feature below the HH band; these differences could be from the inhomogeneous lateral dimensions of the NPL$_A$ batch (**Figure S2**). When the films were excited at low fluences (ca. 0.8 μJ/cm$^2$) and at 420 nm (1 kHz, 100 fs), the photoluminescence showed a single emission band for CdSe NPL$_B$ films but also a weak 5-ML emission peak for CdSe NPL$_A$ films. Although the percentage of 5-ML in CdSe NPL$_A$ is small, fluorescence resonance energy transfer within a 4 ML and 5 ML mixture can convolute carrier lifetime measurements of strongly coupled systems.[49] Also, similar sample heterogeneities are known to result in energy funneling from large to small band gap semiconductors that can result in higher thresholds for amplified spontaneous emission and lasing. Therefore, we use the improved CdSe NPL$_B$ (referred to hereafter as NPL) materials for the strong coupling and polariton lasing study that follows.

To realize strong coupling with our optimized 4-ML CdSe NPLs, we designed Al nanoparticle lattices with varying periods ($a_0$) that support Γ-point ($k_\parallel = 0$) SLR modes above ($a_0 = 270$ nm), between ($a_0 = 310$ nm), and below ($a_0 = 330$ nm) the LH and HH exciton energies. **Figure 1c** shows the angle-resolved transmission of Al lattices for $a_0 = 310$ and 330 nm; data for $a_0 = 270$ nm is in **Figure S3**. Transmission dispersion measurements for bare Al nanoparticle lattices under transverse-electric (TE)-polarized white light exhibit linear dispersive bands that follow the (±1,0) diffraction orders. Both bright and dark modes are observed about the Γ-point where the bands cross, which can be approximated by the Bragg diffraction equation $\lambda = a_0 \times n$. For lattices with $a_0 = 310$ nm, the band edge occurs at 2.52 eV, between the LH and HH energies, and for $a_0 = 330$ nm, at 2.38 eV, below the HH energy. We note an increase in SLR quality (i.e., narrower linewidths) with increased lattice periodicity because of larger detuning of the Rayleigh anomaly (RA) condition from the localized surface plasmon of the nanoparticles, which were kept fixed in size and hence energy (2.70 eV) for all periodicities (**Figure S5**).[34, 50] The transverse magnetic (TM)-polarized dispersion diagrams have quadratic bands and show trends of linewidth narrowing and band-edge lowering similar to that under TE-polarization (**Figure S4**).

**Figure 2** compares experimental and calculated dispersion diagrams of $a_0 = 310$ and 330 nm Al nanoparticle lattices strongly coupled to CdSe NPL films under TE- and TM-polarized light. For both lattice periodicities, the transmission bands are markedly different from bare, uncoupled lattices in Figure 1c and exhibit the characteristic band anti-crossing and band bending of polariton modes. Polariton bands with energies lower than the HH band are identified as lower polaritons (LP); the middle polaritons (MP) are between the HH and LH energies; and the upper polaritons (UP) reside above the LH energy. For the $a_0$ = 310 nm lattice (**Figure 2a**), the TE-polarized transmission shows two distinct LP band behaviors above and below the bandgap (ca. 2.36 eV): (1) a nearly flat-band that terminates at the dark edge of the band near the Γ-point and (2) linear bands that start at the bright edge. The MP and UP bands are relatively flat at large $k_\parallel$ and become more dispersive as they approach $k_\parallel = 0$. Similarly, three polariton bands are experimentally observed for $a_0$ = 330 nm lattices (**Figure 2b**). The LP bands shift to lower energies and become more intense, which makes the dark band-edge more visible. These measurements indicate how a variety of distinct polariton dispersion characteristics can be achieved by tuning the Γ-point SLR energies with respect to the CdSe NPL exciton bands.

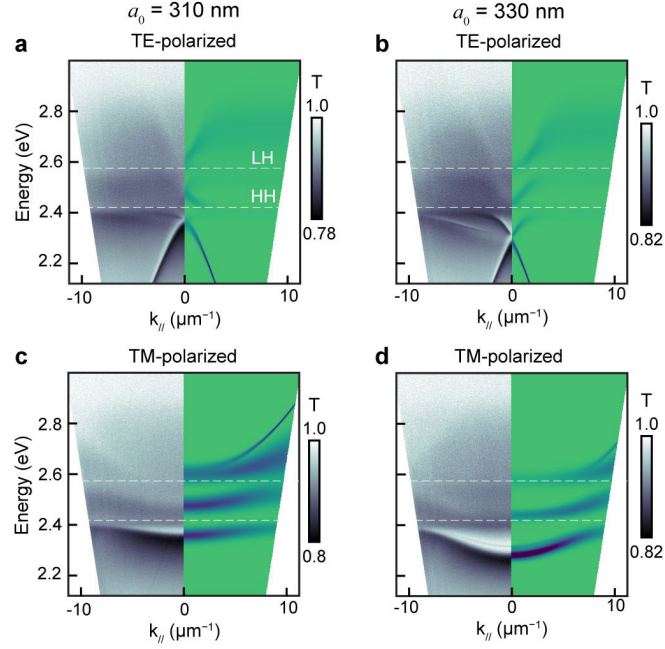

**Figure 2: Polarization-dependent polariton dispersion diagrams.** Experiment (left) and Jaynes-Cummings simulation (right) optical transmission of CdSe NPL films coupled to Al nanoparticle lattices under TE-polarized light with **(a)** $a_0$ = 310 nm and **(b)** $a_0$ = 330 nm and under TM-polarized light with **(c)** $a_0$ = 310 nm and **(d)** $a_0$ = 330 nm. The dashed lines indicate the energies of the light-hole (LH) and heavy-hole (HH) exciton bands.

To confirm that our system is in the strong coupling regime, we used the Jaynes-Cummings model to extract Rabi splitting values ($\Omega$) (**Figure S6-7**).[18] When compared with the exciton ($\gamma_{HH}$, $\gamma_{LH}$,) and SLR ($\gamma_{SLR}$) loss factors (full-width half-maximum (FWHM) of the uncoupled exciton bands and SLR band at the anti-crossing position) (**Table S1**), the strong coupling conditions $\Omega > \gamma_{SLR}$ and $\Omega > \gamma_{HH,LH}$ are satisfied.[7] The right half of each plot in Figure 2 shows the simulated transmission diagrams with linewidths and relative intensities of the polariton bands based on their Hopfield coefficients and uncoupled SLR eigenmode properties.[51, 52] The modelling is in good agreement with experiment; slight discrepancies may be attributed to approximations intrinsic to this model (**Supporting Information: S6-7**).

Since SLR modes are polarization-dependent, strong coupling resulted in polariton bands with distinct dispersion behavior under TM-polarized (versus TE-polarized) light (**Figures 2c-d**). TM-polariton bands have local minima at $k_\parallel = 0$ and show increasingly dispersive character at larger $k_\parallel$ values. As the lattice periodicity $a_0$ of the cavity increased, the polariton bands became narrower while maintaining their flat-banded dispersion (Figure S3, Figures 2a-b). One important feature of the strongly coupled $a_0 = 330$ nm lattice under TM-illumination is the presence of two LP bands (Figure 2d): a slightly blue-shifted, LP band with higher quality factor (Q) and a more intense LP band of lower Q. These LP bands formed from the RA and SLR cavity modes of the plasmonic lattice (**Figure S4**). That there are two LP bands with distinct Q in experiment and theory indicates that both RA and SLR modes can be simultaneously and independently strongly coupled to CdSe NPL excitons. To distinguish between polaritons from RA and SLR modes, we will label them as $LP_{RA}$ and $LP_{SLR}$, respectively.

To characterize the emission of the strongly coupled system, we performed angle-resolved photoluminescence experiments using a Fourier microscopy setup with a 1 kHz, 100 fs pulsed excitation source centered at 420 nm. Surprisingly, lasing only resulted from the $a_0$ = 330 nm lattice that showed the largest negative detuning from the HH exciton, even though the other cavity periodicities also supported strong coupling. **Figure 3a** shows a non-linear increase of the photoluminescence as a function of excitation fluence characteristic of lasing action. At low pump fluence (<5.5 µJ/cm²), only a broad photoluminescence feature from uncoupled NPL excitons was observed around 2.42 eV; however, upon increased fluence, an intense and narrow lasing peak at higher energies ca. 2.35 eV (FWHM = 1.8 meV, 0.4 nm) emerged. The inset in Figure 3 indicates that the lasing emission is delocalized across the entire pump areas, a common feature of plasmonic lattice cavity lasers.[53] Pump fluence vs. output intensity curves show behavior expected of lasing action, with a transition from linear to superlinear behavior accompanied by a rapid narrowing in linewidth (**Figure 3b**). Notably, the threshold fluence $P_{th}$ = 6.0 µJ/cm$^2$ is very low; compared to photon lasing using core-only CdSe NPLs ($P_{th}$ ≈150-200 µJ/cm$^2$),[27, 28] this threshold is around 25-fold less. Despite higher optical losses intrinsic to plasmonic cavities, this low threshold is similar to dielectric-cavity lasers that integrate gain-engineered materials such as core/crown CdSe NPLs or lead halide perovskite nanocrystals.[54-56]

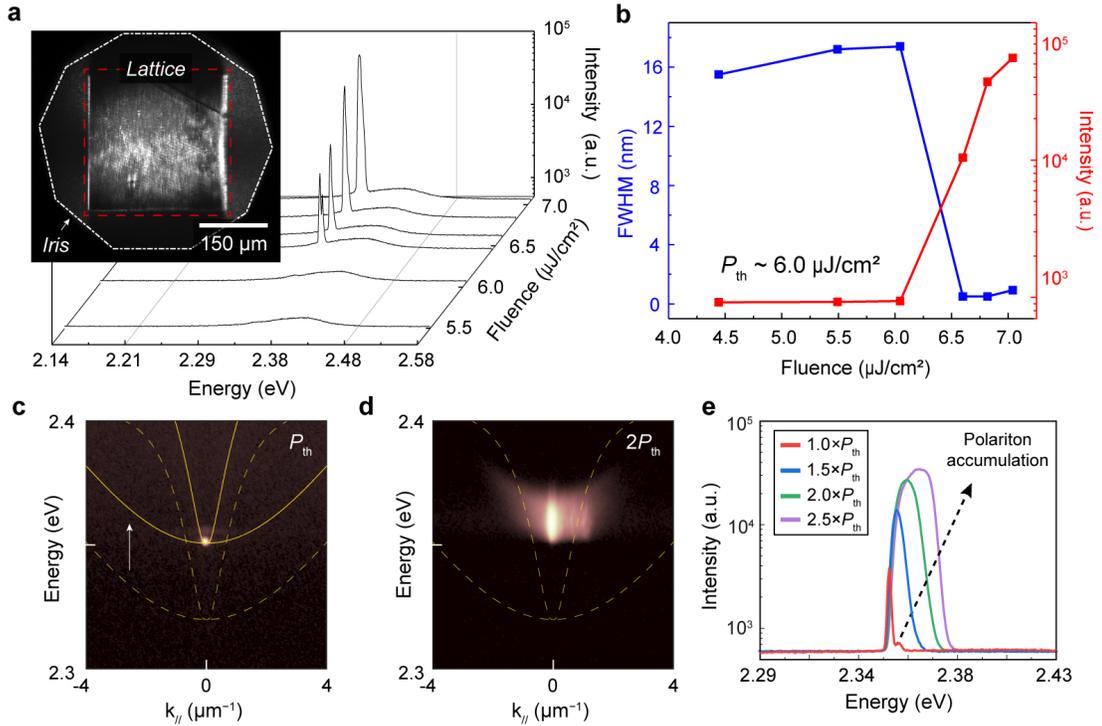

**Figure 3: Polariton lasing from $a_0$ = 330 nm lattice. (a)** Waterfall plot of emission intensity and wavelength as a function of pump fluence. Inset shows the real space image of lattice above lasing threshold $P_{th}$. **(b)** Input-output plot showing a linear to super-linear increase in intensity along with rapid linewidth narrowing at $P_{th}$ ~ 6 µJ/cm². **(c)** Angle-resolved photoluminescence image at $P_{th}$ shows that lasing is slightly blue-shifted at the lasing threshold. LP bands above (below) threshold are indicated in solid (dashed) lines. **(d)** Angle-resolved photoluminescence image at 12 µJ/cm² reveals band broadening and further blue shifting above threshold of the LP bands. **(e)** Above-threshold spectral traces taken at the Γ-point confirm a gradual peak shift and broadening as expected from polariton-polariton interactions.

To determine whether lasing was mediated by photons or polaritons, we used angle-resolved photoluminescence imaging to identify which bands contributed to the signal. At fluences near the threshold (**Figure 3c**), the LP emission occurs at the Γ-point and at an energy 25 meV higher than the isoenergetic TE-polarized dark band-edge SLR and TM-polarized LP$_{RA}$ band (Figures 2b,d). The collapse of emission into the minima of the LP band suggests polariton-polariton relaxation and also that population accumulation at the minima is the mechanism behind the lasing observed, i.e., polariton lasing.[9] At twice the polariton lasing threshold, **Figure 3d** reveals that the emission broadens and shifts to higher energies. Normalized high-fluence spectral measurements (**Figure 3e**) confirm the gradual shift and broadening of the emission from 2.351 eV (FWHM = 1.8 meV, 0.4 nm) at $P_{th}$ to 2.365 eV (FWHM = 14 meV, 3.5 nm) at 2.5 × $P_{th}$. Both the spectral blue-shift and peak broadening are from polariton-polariton Coulombic repulsion and polariton-exciton repulsion.[57] We note that the observed shift in lasing energy is distinct from multi-exciton amplified spontaneous emission from uncoupled CdSe NPLs that show spectral red-shifts due to attractive multi-exciton interactions.[27]

To understand why polariton lasing was only observed from the $a_0 = 330$ nm lattice, we characterized the photoluminescence of the strongly coupled system below threshold (**Figure 4**). Since single-exciton emission in CdSe NPLs occurs as a band that is Stokes-shifted ~7 meV from the HH absorption energy,[27, 48] photoluminescence from polaritons will be limited primarily to LP bands and, in some cases, MP bands; emission from MP is fairly strong for $a_0 = 270$ nm (Figure S3). Although the transmission of the LP bands shows dark and bright SLR band-edges negatively detuned from the HH band for both lattices (Figure 2), the photoluminescence shows additional differences. For the $a_0 = 310$ nm lattice, strong emission is observed from the LP bands, with higher intensities around the Γ-point region, as well as weaker emission from the MP (**Figure 4a, inset**). Since the LP band-edges for the $a_0 = 330$ nm lattice are at lower energies than the HH, the regions of highest intensity are when $k_\parallel > 5$ μm$^{-1}$, particularly for TM-polarized LP bands (**Figure**

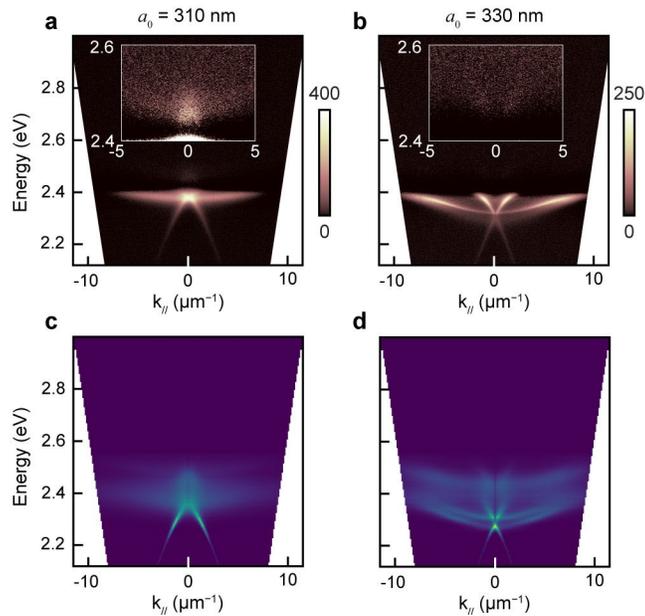

**Figure 4: Cavity-dependent photoluminescence from the lower polariton.** Experimental photoluminescence measurements of Al NP lattices coupled to CdSe NPL films with (**a**) $a_0 =$ 310 nm and (**b**) $a_0 = 330$ nm. Inset in (a) and (b) are regions of middle polariton emission, displayed from 0 to 25 counts. These measurements are background subtracted against the uncoupled film. Simulated PL measurements from a generalized Holstein-Tavis-Cummings model with (**c**) $a_0 = 310$ nm and (**d**) $a_0 = 330$ nm.

**4b**); only faint emission from MPs was measured (inset). Also, we calculated the photoluminescence using the generalized Holstein-Tavis-Cummings (GHTC) model[58] (**Figures 4c-d**), where the population dynamics were simulated using the Lindblad mean-field Ehrenfest method.[59] The photoluminescence spectra were generated by combining polariton population dynamics and Hopfield coefficients of the GHTC exciton-polariton states. The intensities at each angle and energy were weighted by the polariton population and magnitude of SLR character in each state (**Supporting Information: S8-10**). The model correctly predicted emission from the MP and LP bands, in agreement with experiments. Some minor discrepancies are found since the mean-field Ehrenfest dynamics method is known to overestimate the MP and UP populations when using a Holstein phonon-coupling model.[30]

The photoluminescence intensity distribution trends in Figure 4 indicate that cavities with different periodicities facilitate different polariton relaxation pathways. For the $a_0 = 310$ nm lattice, the build-up of the LP population occurs at the bright, TE-polarized band-edge, and then polaritons can decay further along the LP band, which likely precludes polariton accumulation and therefore lasing. In contrast, for the $a_0 = 330$ nm lattice, the LP bands are populated at $k_\parallel > 5$ μm$^{-1}$, and polaritons can lower their energy through polariton-polariton scattering events to accumulate at the band minima. Another determining factor for polariton lasing is the polariton radiative lifetime, which sets an upper limit for the rate that the polariton population can accumulate at the band minima and hence contribute to lasing. Fortunately, the high purity of our 4-ML CdSe NPLs allowed us to estimate uncoupled exciton and LP radiative lifetimes by performing time-resolved photoluminescence that integrated the entire (E, $k_\parallel$) response (Figures 4a-b) using a streak camera under non-resonant, pulsed excitation (420 nm, 0.8 μJ/cm$^2$) (**Supporting Information: S11**). We analyzed time-resolved traces of the LP emission energies below the single-exciton emission peak to extract LP lifetimes.

Radiative decay of uncoupled CdSe NPL films follows a biexponential behavior with room-temperature time constants $\tau_1$ and $\tau_2$ on the order of 30-60 ps and 0.2-12 ns, respectively; the fast and slow decay components are attributed to the bright and dark exciton lifetimes.[16, 17, 45] For the uncoupled NPL film at 2.36 eV, our fitting produced $\tau_1$ = 36.7 ps and $\tau_2$ = 338.6 ps, in agreement with reported values.[16, 17] The strongly coupled system also showed similar biexponential decays, where $\tau_1$ can be assigned to a direct polariton radiative transition and $\tau_2$ to transitions into the longer-lived dark polariton reservoir. Interestingly, the results for $a_0$ = 310 nm lattices ($\tau_1$ = 32 ps and $\tau_2$ = 313.3 ps) showed shorter radiative lifetimes than the uncoupled film, while the $a_0$ = 330 nm lattices ($\tau_1$ = 37.6 ps and $\tau_2$ = 319.8 ps) had longer lifetimes, which favors population accumulation (**Supporting Information: S11**). Photoluminescence and lifetime data suggest that for $a_0$ = 330 nm lattices, the LP bands facilitate efficient polariton relaxation and accumulation at the band minima, where a polariton population involved in lasing can build up because of their longer radiative lifetime. Similarly, polariton accumulation at the bright band-edge of the $a_0$ = 310 nm lattice and the shorter polariton lifetime may have hindered polariton accumulation and, therefore, polariton lasing.

**CONCLUSIONS**

In summary, we demonstrated polariton lasing by tailoring dispersion diagrams of plasmonic lattices to support polariton relaxation and accumulation and by improving the monodispersity of the CdSe NPL gain materials. Interestingly, this lattice cavity architecture can exhibit a rich range of polarization-dependent optical modes for strong coupling, including RA modes and SLRs. Our results highlight prospects for advances in polariton lasing by further engineering of particle shape and lattice geometry. Moreover, unlike the common Fabry-Pérot cavities, plasmonic lattices have an open architecture that allows ready interfacing with microfluidic devices as well as optical and electrochemical platforms to study polariton chemistry. We anticipate that ease in tuning polariton bands can be leveraged to modify the potential energy landscape of chemical reactions to increase yield or achieve unexpected products, opening new tools for synthetic chemists.

**METHODS**

**Sample fabrication**

Arrays of Al NPs were fabricated using electron-beam lithography (Raith Voyager) on PMMA A3 (Kayaku Advanced Materials) coated H-BAK8 glass substrates (CDGM). After development, 55 nm of Al were electron-beam evaporated on the patterned substrates and then the photoresist was lifted off using acetone. 4 monolayer CdSe core nanoplatelet films were spin cast onto Al NP arrays by dropping 45 µL of a 60 mg/mL NPL solution while the substrate was spinning at 4000 rpm. The full synthesis and purification protocols of the CdSe NPLs can be found in the Supporting Information.

**Optical characterization**

Wavelength and angle-resolved transmission and PL measurements were carried out using a Fourier microscopy setup. The back focal plane of a 20x Plan Apo Nikon (NA = 0.75) objective lens was imaged onto the entrance slit of a SP2500 (Teledyne, Princeton Instruments) spectrometer coupled to a 2D CCD camera (PIXIS 400, Teledyne, Princeton Instruments). The images are then converted into (E, $k_\parallel$) plots where E = $hc/\lambda$ and $k_\parallel$ = $(\omega/c)\sin\theta$ where $h$ is the Planck constant, $c$ is the speed of light in vacuum, $\lambda$ is the wavelength, $\omega$ is the angular frequency, and $\theta$ is the incidence angle. During the transmission measurements, the samples were illuminated using a broadband white-light source whose polarization was controlled to measure the optical response under TE or TM-polarized excitation.

During the PL and lasing experiments, the sample was excited at 420 nm by a laser system consisting of an tunable optical parameter amplifier (TOPAS Prime, Light Conversion) and a Solstice Ace (Spectra-Physics) mode-locked Ti:sapphire laser with a regenerative amplifier (fundamental 800-nm wavelength, 1000-Hz operation, and 100-fs pulse width). The laser excitation was incident at ca. 60 deg with respect to the sample normal and had a spot size of 3.45 mm by 1.7 mm. To remove the signal from uncoupled emitters in Figure 4, the PL was background subtracted against a pristine CdSe NPL film region located beside each array which minimized film heterogeneity effects. In the inset in Figure 3a, the pump beam spot size is larger than the imaged area and filtered out during imaging using a longpass filter.

Time-resolved photoluminescence measurements were performed using a streak camera (Hamamatsu) system with an Acton SP150 spectrometer (Princeton Instruments). During the experiments, the samples were excited using at 420 nm with a fluence of 0.8 μJ/cm$^2$ with the same laser system used in the lasing experiments. To image the full angular range capture by our Fourier microscopy setup, the back focal plane image of the collection objective lens was image onto the entrance slit of the spectrometer.

**ACKNOWLEDGEMENTS**

This work was supported by the National Science Foundation (NSF) Center for Quantum Electrodynamics for Selective Transformations (QuEST) (Grant CHE-2124398, N.G.S. E.K, S.-M.P., P.H., T.D.K.), and the Vannevar Bush Faculty Fellowship from the U.S. Department of Defense (DOD N00014-17-1-3023, F.F.F., M.J.H.T., T.W.O.). This work made use of the NUFAB, EPIC, and SPID facilities of Northwestern University's NUANCE Center that has received support from the SHyNE Resource (NSF ECCS-2025633), the International Institute for Nanotechnology at Northwestern University (IIN), Northwestern's MRSEC Program (NSF DMR-1720139), and MRI (NSF DMR-1828676). M.J.T. and S.-M.P. gratefully acknowledge support from the Ryan Fellowship and the IIN. This research was supported in part by the Quest high performance computing facility at Northwestern University, which is jointly supported by the Office of the Provost, the Office for Research, and Northwestern University Information Technology. Computing resources were also provided by the Center for Integrated Research Computing (CIRC) at the University of Rochester. We also acknowledge Lele Mathis, Dr. Alexander D. Sample and Dr. Xitlali G. Juarez at Northwestern University for initial assistance with sample fabrication and discussion and William Girten and Dr. Nicole M.B. Cogan at University of Rochester for initial assistance with NPL synthesis and sample characterization.

**ASSOCIATED CONTENT**

Supporting Information is available free of charge at XXX

- CdSe NPL synthesis and purification; NPL and film characterization; Additional optical characterization; Localized Surface Plasmon mode simulation; Details of Fitting Parameters; Theoretical Details of the SLR-Mode Dispersion; Theoretical Details of the $k_\parallel$-resolved Transmission Spectra Simulations; Theoretical Details of the $k_\parallel$-resolved Photoluminescence Spectra Simulations; Details of the Exciton-Phonon Coupling Hamiltonian; Details of the Quantum Dynamics Simulations; Streak Camera Measurements and Fittings